\documentclass[prx,twocolumn,floatfix,superscriptaddress,aps,longbibliography]{revtex4-2} %,linenumbers

\usepackage{amsfonts,amsmath,amssymb,color,times,graphicx}
\definecolor{darkblue}{rgb}{0, 0, 0.8}
\usepackage[colorlinks=true, breaklinks=true, linkcolor=darkblue, citecolor=darkblue, urlcolor=darkblue]{hyperref}

\usepackage{graphicx}
\usepackage{bmpsize}
\usepackage{amsfonts}
\usepackage{gensymb}
\usepackage{braket}
\usepackage{mathtools}
\usepackage{bm}
\usepackage{color}

\begin{document}
\title{Helical edge states and enhanced superconducting gaps in Bi islands on FeTe$_{0.55}$Se$_{0.45}$}

\author{Chuanhao Wen, Zhiyong Hou, Zhiyuan Shang, Huan Yang$^*$, and Hai-Hu Wen$^\dag$}

\affiliation{National Laboratory of Solid State Microstructures and Department of Physics, Collaborative Innovation Center of Advanced Microstructures, Nanjing University, Nanjing 210093, China}

\begin{abstract}
 By measuring scanning tunneling spectroscopy on some large Bi islands deposited on FeTe$_{0.55}$Se$_{0.45}$ superconductors, we observe clear evidence of topological in-gap edge states with double peaks at about $\pm 1.0$ meV on the spectra measured near the perimeter of the islands. The edge states spread towards the inner side of the islands over a width of 2-3 nm. The two edge-state peaks at positive and negative energies both move to higher values with increase of the magnetic field, and they disappear near the transition temperature $T_\mathrm{c}$ of FeTe$_{0.55}$Se$_{0.45}$. The edge states are interpreted as the counter-propagating topological edge states induced by the strong spin-orbit coupling effect of the Bi island, and the zero-energy mode emerges when the edge states touch each other in some small Bi islands. Meanwhile, enhanced superconducting gaps are observed in the central regions of these Bi islands, which may be induced by the enhanced pair potential of the topological surface state. Our observations provide useful message for the nontrivial topological superconductivity on specific Bi islands grown on FeTe$_{0.55}$Se$_{0.45}$ substrate.
\end{abstract}

\maketitle

Topological materials and related physics have attracted significant attention in the field of condensed matter physics \cite{RN001,RN002}. One of the most fundamental properties of these topological materials is the unique feature at the boundary, for example there are gapless states on the surface of a three-dimensional (3D) bulk, or edge states at the edge of two-dimensional (2D) material, and even at the two terminals of one dimensional bar. As a prominent representative, the 3D topological insulators, with insulating gaps in the bulk, have features of metallic topological Dirac surface states. Experimentally, topological boundary states were observed in HgTe quantum wells \cite{RN003,RN004}, Bi$_x$Sb$_{1-x}$ \cite{RN005,RN006}, Bi$_2$Se$_3$ \cite{RN007,RN008}, Bi$_2$Te$_3$ \cite{RN009}, Sb$_2$Te$_3$ \cite{RN010} crystals. In the 2D case, two helical edge states with opposite spins may propagate in opposite directions at the edges \cite{RN011}. Similar to topological insulators, topological superconductors also possess topologically protected gapless states at the boundary, which is regarded as the hallmark for the existence of the Majorana fermions. In previous studies, zero-bias conductance peaks (ZBCPs) were observed at the terminals of chains and the edges of two-dimensional islands adjacent to superconductors \cite{RN012,RN013,RN014,RN015,RN016} by using scanning tunneling microscopy/spectroscopy (STM/STS). Similar ZBCPs were also observed in vortex cores on the surfaces of topological insulators on top of some iron-based superconductors \cite{RN017,RN018,RN019,RN020,RN021,RN022,RN023}. Almost all these ZBCPs were attributed to Majorana zero modes (MZMs), which hold great promise for future applications in quantum computing devices.

As a heavy element with strong spin-orbit coupling, bismuth plays a crucial role in constructing topological states in some materials. Apart from being a key element in many topological materials like Bi$_x$Sb$_{1-x}$, Bi$_2$Te$_3$ and PbBi$_2$Te$_4$, Bi itself exhibits very fascinating electronic states. Previous experiments identified the presence of surface states with a topologically nontrivial origin in the bulk Bi \cite{RN024,RN025,RN026}, and the ultrathin Bi film is theoretically predicted to be a 2D topological insulator \cite{RN027,RN028}. The topological edge states were observed on one to dozens of bilayers (BLs) of Bi(111) and Bi(110) grown on various substrates \cite{RN029,RN030,RN031,RN032,RN033,RN034,RN035,RN036}. Since the edge states are protected by time-reversal symmetry, the Bi films may be ideal platforms for MZMs when superconductivity is induced by the proximity effect from a neighboring superconductor. This scenario has been realized on the interface between the iron cluster and the helical edge state of Bi(111) island grown on a superconducting niobium substrate \cite{RN036}. In our previous studies, ZBCPs were observed on some Bi islands with a round shape and areas of 10–50 nm$^2$ deposited on the superconducting substrate Fe(Te,Se) \cite{RN037,RN038}. However, for larger islands, it remains to be answered whether this nontrivial state still exists. Thus, it is very essential to prove the existence of the topological edge modes in the Bi islands on Fe(Te,Se) with larger size and irregular shapes, which may unify and enhance our understanding on proximity induced topological superconductivity in this heterostructure.

\begin{figure*}[tb]
 \centering
 \includegraphics[width=17cm]{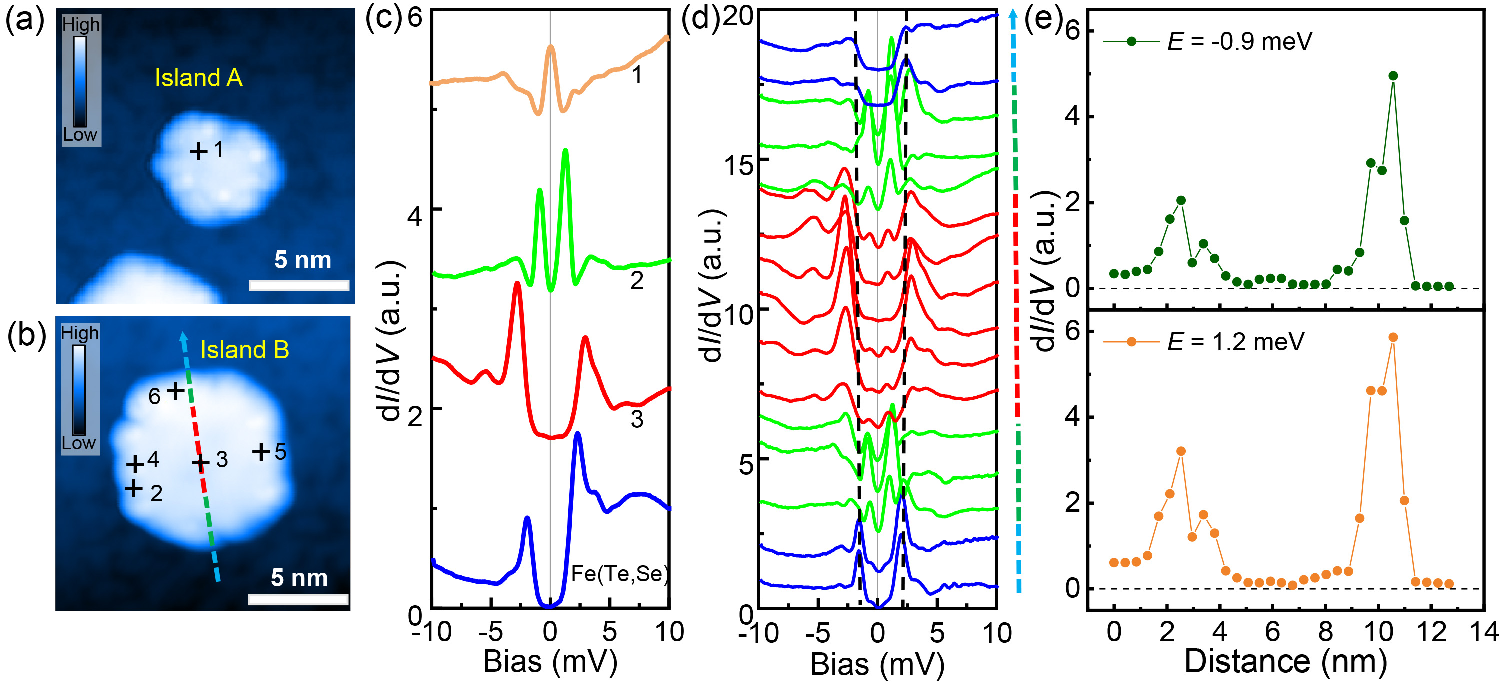}
 \caption{(a),(b) Topographic images of the Bi islands A and B on the surface of the Fe(Te,Se) (setpoint conditions: $V_\mathrm{set}= 1$ V, $I_\mathrm{set}= 20$ pA). (c) Tunneling spectra measured on selected points on the islands and Fe(Te,Se) substrate nearby ($V_\mathrm{set}= 10$ mV, $I_\mathrm{set}= 200$ pA). The spectra have been vertically offset for clarity. (d) A set of tunneling spectra measured along the arrowed dashed line in (b) crossing the island B ($V_\mathrm{set}= 10$ mV, $I_\mathrm{set}= 200$ pA). The curves are plotted in the same color as the corresponding arrow segment in (b). The green segments correspond to the region with the existence of the edge state. (e) Spatial dependence of d$I$/d$V$ obtained from all the spectra measured along the arrowed line in (b), and the selected energies correspond to the in-gap edge states at negative and positive energies, i.e., $-0.9$ and $1.2$ meV.}
 \label{fig.1}
\end{figure*}

In this Letter, we report the observation of topological edge states on Bi islands with the lateral size larger than 8 nm deposited on the iron-based superconductor FeTe$_{0.55}$Se$_{0.45}$. These edge states exhibit a double-peak feature on the STS, which spreads over a width of 2–3 nm at the entire perimeter of the island. As the tip moves away from the edge, the low energy peaks of tunneling spectra rapidly decay and become almost invisible in the central region. The energy positions of these edge state peaks shift linearly to the higher energy with increasing magnetic fields. Another interesting finding on these islands is that the induced superconducting gap is significantly enhanced as a result of parity mixing of pair potential. As the temperature increases, both the edge states and enhanced gaps vanish almost simultaneously with the gap-closing temperature of the substrate Fe(Te,Se). Our results reveal that the topological nontrivial superconductivity has been induced on these islands.

Bi islands were grown and investigated in a USM-1300 system (Unisoku Co. Ltd.) with a molecular beam epitaxy chamber, and details about the experimental methods are described in the Supplemental Material \cite{RN039}. The obtained Bi islands randomly distribute on the surface of the Fe(Te,Se) substrate. These islands have the dimensions of several nanometers and heights of about 7 \AA\, and the atomic structures of most islands are distorted orthorhombic with lattice parameters of about 4.5-4.8 \AA\ \cite{RN037,RN038}. These parameters suggest that most Bi islands consist of Bi(110) bilayer films \cite{RN037,RN038,RN035,RN040,RN041,RN042}. From our previous works \cite{RN037,RN038}, the ZBCPs exist on some of the Bi islands with a diameter of 4-8 nm. For example, the topography of a Bi island (island A) holding the ZBCP is shown in Fig.~\ref{fig.1}(a), and the diameter of the island is about 5 nm. A typical tunneling spectrum measured on point No. 1 is shown in Fig.~\ref{fig.1}(c), and one can see clear ZBCP on the spectrum. This is different from the typical superconducting spectrum measured on the Fe(Te,Se) substrate nearby, which behaves as the U shape near zero bias, denoting a fully gapped feature. However, the ZBCP disappears when the area of the island exceeds about 50 nm$^2$ \cite{RN038}. The topography of such an island (island B) is shown in Fig.~\ref{fig.1}(b), and the diameter of this island is about 9 nm. The surface structure of the island is also a distorted orthorhombic one. In addition to the height of about 7 \AA, we argue that the Bi island reported here also consists of Bi(110) bilayer film. The tunneling spectrum measured near the center of this island (point No. 3) is shown in Fig.~\ref{fig.1}(c), and the ZBCP is absent on the spectrum, which is consistent with our previous conclusion of the absence of the ZBCP in the island with the area exceeding about 50 nm$^2$ \cite{RN038}. However, on the spectrum measured near the island edge (point No. 2), there are two sharp peaks within the coherence-peak energy. These two peaks are at about $-0.9$ and $1.2$ meV, respectively, with slight particle-hole asymmetry.

To further investigate the in-gap edge states, we measure a sequence of tunneling spectra along the arrowed line in Fig.~\ref{fig.1}(b) across island B, and the results are shown in Fig.~\ref{fig.1}(d). The edge-state peaks near $-0.9$ and $1.2$ meV can be clearly observed on spectra measured at positions within about 2 nm from the island edge contour, and the edge-state peaks weaken when the tip moves to the center of the Bi island. In Fig.~\ref{fig.1}(e) , we plot the spatial dependence of the differential conductance at edge-state energies, i.e., $-0.9$ and $1.2$ mV. One can see that the edge state mainly exists in the region of 2-3 nm from the island edge. As the tip moves inwards, the edge state diminishes and becomes almost indistinguishable at the center of the island. The width of the edge state in present work aligns with the spatial scales at which topological edge states are found on Bi islands grown on other substrates \cite{RN029,RN030,RN035}. Besides the asymmetric edge-state energies, the heights for the peaks at positive and negative energies are also different, and the peak height on the negative-energy side is always lower than that on the positive-energy side. 

The coherence peaks on the tunneling spectrum measured in a superconductor usually reflect the superconducting gap value or the gap maximum. For example, in Fig.~\ref{fig.1}(d), the spectra measured on the Fe(Te,Se) substrate show coherence peaks at about $\pm 2.2$ mV which are indicated by the vertical dashed lines. In the center region of the Bi island, the in-gap edge states significantly weaken. Meanwhile, the coherence peaks move to the higher energies at about $\pm 3.0$ mV on these spectra, which has a 36\% enhancement of the superconducting gap of the Fe(Te,Se) substrate. In addition, the coherence peaks become pronounced on these spectra, which suggests an enhancement of the superconductivity. The enhanced superconducting gap is also observed on the spectra measured in some large Bi islands without the in-gap edge states or the ZBCP from our previous work \cite{RN038}.

\begin{figure}[tb]
  \includegraphics[width=\columnwidth]{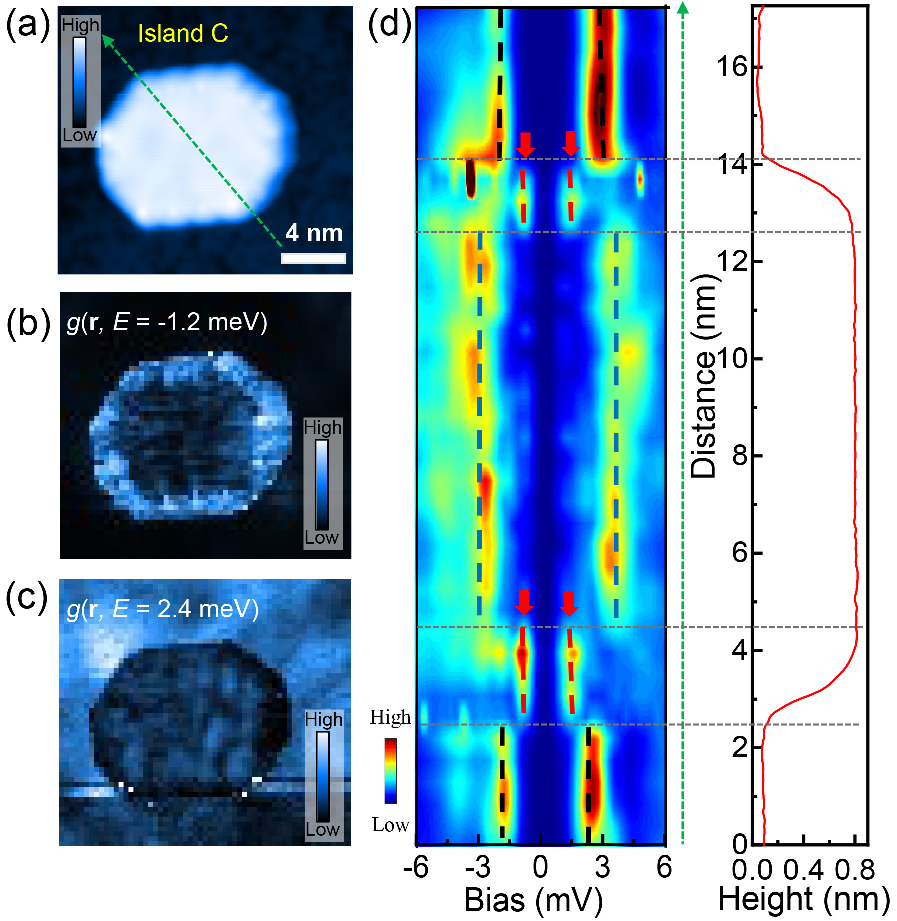}
  \caption{(a) Topographic image of the Bi island C ($V_\mathrm{set}= 1$ V, $I_\mathrm{set}= 20$ pA). (b),(c) Differential conductance mappings measured in the area of (a) at the biases of $-1.2$ and $2.4$ mV, respectively ($V_\mathrm{set}= 10$ mV, $I_\mathrm{set}= 50$ pA). (d) Left panel: Color plots of a set of tunneling spectra measured along the arrowed dashed line in (a) ($V_\mathrm{set}= 10$ mV, $I_\mathrm{set}= 200$ pA). The red, black, and blue vertical dashed lines mark the averaged peak positions of edge-state peaks, coherence peaks of the spectra measured on the Fe(Te,Se) substrate, and coherence peaks of the spectra measured in the central area of the Bi island, respectively. Right panel: Height profile along the arrowed dashed line in (a). The gray horizontal dotted lines mark the spatial range where edge states appear.}
  \label{fig.2}
\end{figure}

As mentioned in our previous works \cite{RN037,RN038}, it is difficult to grow Bi islands with very large size on Fe(Te,Se). Consequently, it is rare to see the edge states on these islands. To rule out the possibility for observing the edge states by accident, control experiments are conducted on Bi island C, and the results are shown in Fig.~\ref{fig.2}. Figure~\ref{fig.2}(a) shows the topography of the island, and the area of the island is about 85 nm$^2$. In this island, the in-gap edge state appears at about $-1.2$ and $1.0$ mV. Figure~~\ref{fig.2}(b) shows the differential conductance mapping measured at $-1.2$ mV, and a bright uninterrupted ring with a width of 2-3 nm is observed along the entire perimeter of the Bi island. Since the differential conductance is related to the local density of states, the bright ring is the edge states along the entire boundary of the island. Meanwhile, the width of the bright ring varies along different direction, and the inner boundary is also irregular. This suggests an inhomogenous distribution of the edge state. This inhomogeneity may be attributed to the imperfect assembly of the Bi island. In spite of the ubiquitous disturbance by defects, the edge states exist continuously, indicating that they are topologically protected. The left panel of Fig.~\ref{fig.2}(d) shows the color plot of the tunneling spectra measured along the arrowed line in Fig.~\ref{fig.2}(a). The spectrum features are similar to those obtained in island B. The edge state can be clearly seen in the edge region, and the coherence-peak energy also enhances in the central region of the Bi island than that in the Fe(Te,Se). The averaged gap values in the central area of the Bi island and on the Fe(Te,Se) substrate are $3.1$ and $2.3$ meV, respectively. Figure~\ref{fig.2}(c) shows the differential conductance mapping at $2.4$ meV which is near the gap of Fe(Te,Se). One can see that the d$I$/d$V$ enhances in the Fe(Te,Se) substrate, while that is very low within the Bi island. This is also consistent with the enhanced superconducting gap within the Bi island, especially in the central region. Therefore, the existence of the in-gap edge state and the enhancement of the superconducting gap are well reproduced in the Bi island C. Similar edge states have also been observed in some other islands, as shown in Fig.S3 and Fig.S4\cite{RN039}.

\begin{figure}[tb]
  \includegraphics[width=0.95\columnwidth]{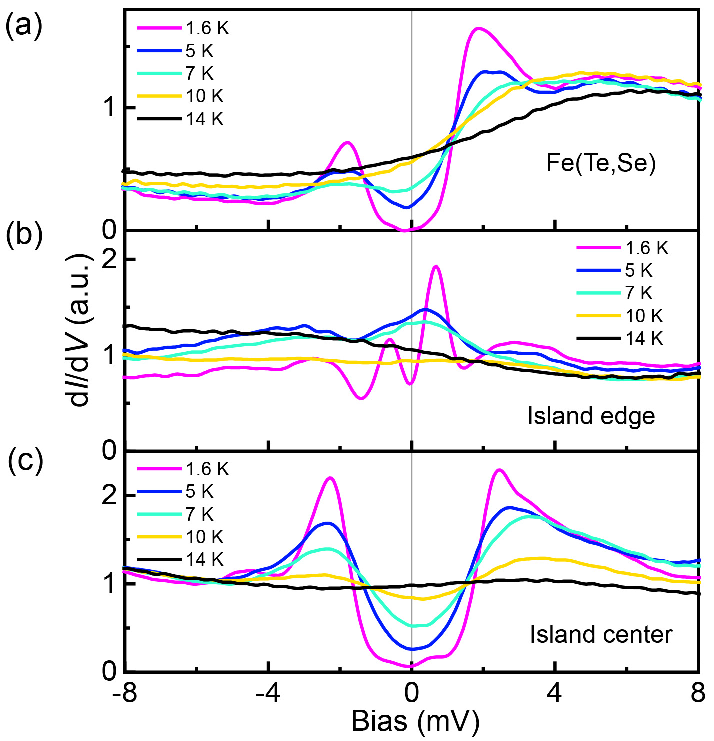}
  \caption{Tunneling spectra measured at different temperatures and in (a) the Fe(Te,Se) substrate, (b) the edge region of the Bi island B, and (c) the central area of this island, respectively ($V_\mathrm{set}= 10$ mV, $I_\mathrm{set}= 200$ pA).}
  \label{fig.3}
\end{figure}

In order to have a further investigation on the edge state and the enhanced superconducting gap, the temperature dependent experiment of the tunneling spectrum is performed, and the results are shown in Fig.~\ref{fig.3}. On the Fe(Te,Se) substrate, the superconducting gapped feature on the spectrum is suppressed with an increase of temperature, and the feature finally disappears at 14 K [Fig.~\ref{fig.3}(a)], which is  the critical temperature of the bulk. Meanwhile, from Fig.~\ref{fig.3}(b), the in-gap edge state in the edge region of the Bi island is also significantly suppressed by increasing temperature, and edge-state peaks disappear at about 10 K, a bit lower than $T_\mathrm{c}$ of the substrate. Since 10 K is not high enough to eliminate the sharp peaks by thermal broadening effect, the edge state is related to the proximity-induced superconductivity on the Bi island. In the central area of the large Bi island, the superconducting gap is significantly enhanced, and the superconducting gapped feature also completely disappears at $T_\mathrm{c}$ of Fe(Te,Se) [Fig.~\ref{fig.3}(c)], although the gap value has a 36\% enhancement in the region. Therefore, $T_\mathrm{c}$ of the central area is not enhanced despite the larger superconducting gap, and there may be an increase in the strength of the superconducting coupling in such area.

\begin{figure}[tb]
  \includegraphics[width=\columnwidth]{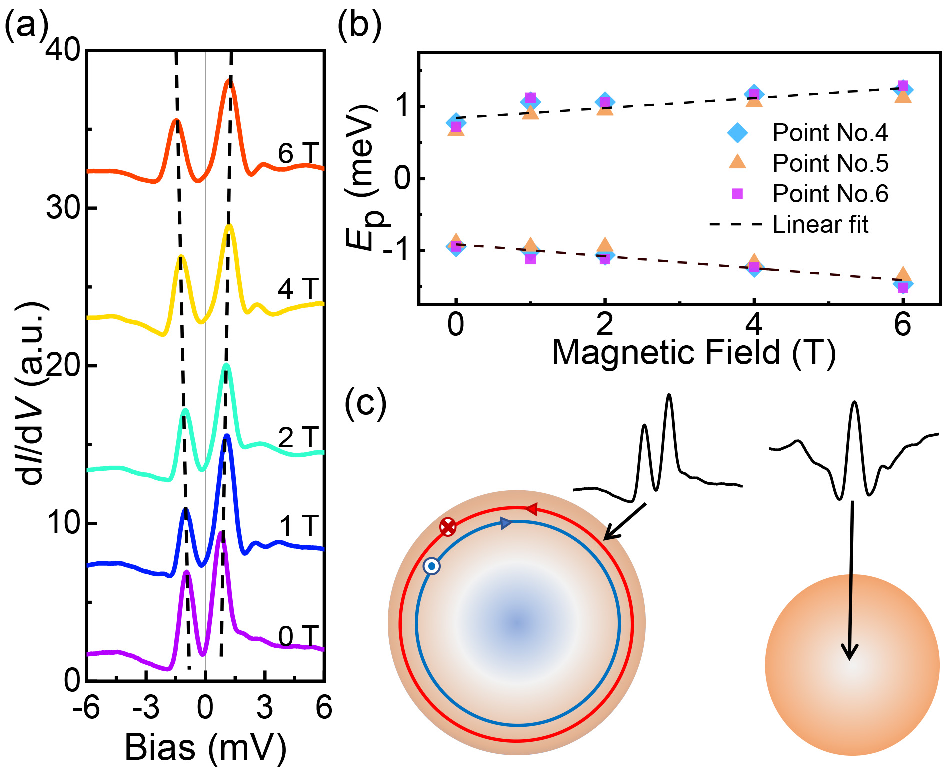}
  \caption{(a) Tunneling spectra measured at different magnetic fields at point No.4 within the edge region shown in Fig.~\ref{fig.1}. The spectra have been vertically offset for clarity. The setting conditions are ($V_\mathrm{set}= 10$ mV, $I_\mathrm{set}= 200$ pA). The dashed lines track the peak positions of edge states. (b) Magnetic field dependent peak positions of the edge states obtained at different places within the island edge. The dashed lines represent the linear fitting results. (c) Sematic images of edge states in Bi islands with different sizes.}
  \label{fig.4}
\end{figure}

The ZBCPs appearing in some small Bi islands show anomalous behaviour under varying magnetic fields from our previous works \cite{RN038}, it is interesting to examine how the in-gap edge-state peak changes at a magnetic field. Figure~\ref{fig.4}(a) shows the field evolution of tunneling spectra measured at point No. 4 marked in Fig.~\ref{fig.1}(b). As can be seen, the edge-state peaks are gradually suppressed with the increase of magnetic field. More importantly, both peaks move to higher energies at higher fields. We extract the energies of the peaks for this point and two others from different edges of island B, and plot them in Fig.~\ref{fig.4}(b). One can see that the absolute values of the edge-state peak energies increase linearly with the increase of magnetic field, and the increasing ratio is about $0.083$ meV/T from the linear fitting. This change suggests that the edge states can be modified by magnetic field. 

As presented above, in some large Bi islands consisting of the Bi(110) bilayer films, we observe strong in-gap edge states at about $\pm 1$ meV within the superconducting gap of Fe(Te,Se) or the proximity-induced superconducting gap on the Bi islands. The edge states exist within about 2 nm from the island edge, and the width is similar to other topological edge states appearing on Bi islands grown on different substrates \cite{RN029,RN030,RN031,RN032,RN033,RN034,RN035,RN036}. These edge states exhibit different features at variable energies, but edge states with an in-gap double-peak feature have not been experimentally observed in heterostructures of bismuth. This in-gap state is schematically shown in the left panel of Fig.~\ref{fig.4}(c). The edge-state peaks disappear near $T_\mathrm{c}$ of the Fe(Te,Se) substrate. In addition, the peaks move to higher energies with an increase of magnetic field. The slope of about $0.083$ meV/T is close to the value of $g\mu_\mathrm{B}=0.12$ meV/T with $\mu_\mathrm{B}$ the Bohr magneton and Land\'{e} $g$-factor $g=2$ for free electrons. These features mentioned above are very similar to the behavior of the Yu-Shiba-Rusinov states induced by the magnetic impurities on superconductors \cite{RN043,RN044}. However, bismuth is a nonmagnetic metal, and it is difficult to have a magnetic impurity ring around the Bi island. Moreover, the edge states observed here spread continuously over the entire perimeter, which contrasts significantly with the localized characteristic of the Yu-Shiba-Rusinov states. Since the conduction electrons of Bi atoms have a strong spin-orbit coupling effect, it is natural for us to assume that there are two counter-propagating topological edge states, and the momentum as well as the spins of these two edge states should have helical feature \cite{RN037,RN038}. These two edge states may result in characteristic peaks within the superconducting gap. The spin property of the counter-propagating topological edge states may result in the field-dependent behavior of the edge-state peak energy \cite{RN045}. Meanwhile, we note that the two in-gap states are theoretically predicted on the surface of 3D topological superconductors in the situation of topological odd-parity gap or some special Fermi surfaces \cite{RN046,RN047}. In the present 2D case, it is very likely that similar boundary state will emerge as the observed edge states. However, to fully understand the in-gap edge states appearing on the large Bi island in this work, further theoretical calculations are highly desired.    

It should be noted that the areas of Bi islands with edge states are all larger than 55 nm$^2$, whereas our previous work \cite{RN037,RN038} show that the areas of islands with ZBCPs mainly range from 10 to 50 nm$^2$. When the islands hosting the edge states are very small with a diameter of 4-8 nm, the edge states may touch each other and form an interfered resonant state as ZBCP \cite{RN038}, which is schematically shown in the right panel of Fig.~\ref{fig.4}(c). Islands with edge states and those with ZBCPs are essentially the same types, and the difference in size produces different phenomena. On some large Bi islands grown on the Fe(Te,Se) substrates with or without \cite{RN038} the edge states, we also observe the enhanced superconducting gaps. The proximity-induced gaps in the conducting part of most heterostructures are usually smaller than that of the superconducting part \cite{RN016,RN019,RN048,RN049,RN050,RN051}. However, the superconducting gap has an enhancement of about 36\% in the large Bi islands, suggesting an enhancement of the superconducting pair potential. We note that the surface superconducting gap is theoretically suggested to be larger than the bulk one due to the parity mixing of the pair potential \cite{RN047}. In the present heterostructure, the surface state may coexist with the topological edge state. This surface state may have an odd-parity superconductivity and a larger superconducting gap. A similar feature of the enhanced superconductor gap has been observed on $\beta$-Bi$_2$Pd thin films grown on SrTiO$_3$ substrate \cite{RN052}.  

In conclusion, we observe topological edge states on the tunneling spectra of some specific Bi islands grown on the iron based superconductor Fe(Te,Se). These edge states exhibit in-gap double peaks and display a 2-3 nm wide spatial distribution near the island perimeter. In the central regions of these islands, enhanced superconducting gaps are obtained, which is attributed to the parity mixing of pair potential induced by surface Dirac fermions. The appearance of edge states and enhanced gaps on the Bi/Fe(Te,Se) heterostructure suggests the non-trivial superconductivity with odd parity on the Bi islands. Moreover, the interaction of the counter-propagating edge states may be the right reason for the presence of ZBCPs on smaller islands \cite{RN037,RN038}. Due to the easy preparation and fine-tuning of these edge states, the recipe for depositing Bi islands on Fe(Te,Se) emerges as a promising candidate for producing the topological superconducting devices for the purpose of quantum computation. 

We appreciate very useful discussions with J. Schmallian and N. Hao. The work was supported by the National Key R\&D Program of China (Grants No. 2022YFA1403201 and No. 2024YFA1408104), and the National Natural Science Foundation of China (Grants No. 11927809 and No. 12434004).

$^*$ huanyang@nju.edu.cn, $^\dag$ hhwen@nju.edu.cn


\begin{thebibliography}{59}

\bibitem{RN001} X. L. Qi and S. C. Zhang, Topological insulators and superconductors, Rev. Mod. Phys. \textbf{83}, 1057 (2011).

\bibitem{RN002} M. Sato and Y. Ando, Topological superconductors: A review, Rep. Prog. Phys. \textbf{80}, 076501 (2017).

\bibitem{RN003} M. König, S. Wiedmann, C. Brüne, L. W. Molenkamp, H. Buhmann,X.-L. Qi, and S.-C. Zhang, Quantum Spin Hall Insulator State in HgTe Quantum Wells, Science. \textbf{318}, 766 (2007).
 
\bibitem{RN004} A. Roth, C. Brüne, H. Buhmann, L. W. Molenkamp, J. Maciejko, X.-L. Qi, and S.-C. Zhang, Nonlocal transport in the quantum spin Hall state, Science \textbf{325}, 294 (2009).

\bibitem{RN005} D. Hsieh, D. Qian, L. Wray, Y. Xia, Y. S. Hor, R. J. Cava, and M. Z. Hasan,  A topological Dirac insulator in a quantum spin Hall phase, Nature (London) \textbf{452}, 970 (2008).

\bibitem{RN006} D. Hsieh, Y. Xia, L. Wray, D. Qian, A. Pal, J. H. Dil, J. Osterwalder, F. Meier, G. Bihlmayer, C. L. Kane, Y. S. Hor, R. J. Cava, and M. Z. Hasan, Observation of unconventional quantum spin textures in topological insulators, Science \textbf{323}, 919 (2009).

\bibitem{RN007} D. Hsieh, Y. Xia, D. Qian, L. Wray, J. H. Dil, F. Meier, J. Osterwalder, L. Patthey, J. G. Checkelsky, N. P. Ong, A. V. Fedorov, H. Lin, A. Bansil, D. Grauer, Y. S. Hor, R. J. Cava, and M. Z. Hasan, tunable topological insulator in the spin helical Dirac transport regime, Nature (London) \textbf{460}, 1101 (2009).

\bibitem{RN008} Y. Xia, D. Qian, D. Hsieh, L. Wray, A. Pal, H. Lin, A. Bansil, D. Grauer, Y. S. Hor, R. J. Cava, and M. Z. Hasan, Observation of a large-gap topological-insulator class with a single Dirac cone on the surface, Nat. Phys. \textbf{5}, 398 (2009).
 
\bibitem{RN009}	Y. L. Chen, J. G. Analytis, J. H. Chu, Z. K. Liu, S. K. Mo, X. L. Qi, H. J. Zhang, D. H. Lu, X. Dai, Z. Fang, S. C. Zhang, I. R. Fisher, Z. Hussain, and Z. X. Shen, Experimental Realization of a Three-Dimensional Topological Insulator, ${\mathrm{Bi}}_{2}{\mathrm{Te}}_{3}$, Science \textbf{325}, 178 (2009).   
    
\bibitem{RN010}	D. Hsieh, Y. Xia, D. Qian, L. Wray, F. Meier, J. H. Dil, J. Osterwalder, L. Patthey, A. V. Fedorov, H. Lin, A. Bansil, D. Grauer, Y. S. Hor, R. J. Cava, and M. Z. Hasan, Observation of Time-Reversal-Protected Single-Dirac-Cone Topological-Insulator States in ${\mathrm{Bi}}_{2}{\mathrm{Te}}_{3}$ and ${\mathrm{Sb}}_{2}{\mathrm{Te}}_{3}$, Phys. Rev. Lett. \textbf{103}, 146401 (2009).
    
\bibitem{RN011} X.-L. Qi, T. L. Hughes, S. Raghu, and S.-C. Zhang, Time-Reversal-Invariant Topological Superconductors and Superfluids in Two and Three Dimensions, Phys. Rev. Lett. \textbf{102}, 187001 (2009).
    
\bibitem{RN012} S. Nadj-Perge, I. K. Drozdov, J. Li, H. Chen, S. Jeon, J. Seo, A. H. MacDonald, B. A. Bernevig, and A. Yazdani, Observation of Majorana fermions in ferromagnetic atomic chains on a superconductor, Science \textbf{346}, 602 (2014).

\bibitem{RN013} G. C. Ménard, S. Guissart, C. Brun, R. T. Leriche, M. Trif, F. Debontridder, D. Demaille, D. Roditchev, P. Simon, and T. Cren, Two-dimensional topological superconductivity in Pb/Co/Si(111), Nat. Commun. \textbf{8}, 2040 (2017).

\bibitem{RN014} R. M. Lutchyn, E. P. A. M. Bakkers, L. P. Kouwenhoven, P. Krogstrup, C. M. Marcus, and Y. Oreg, Majorana zero modes in superconductor–semiconductor heterostructures, Nat. Rev. Mater. \textbf{3}, 52 (2018).
    
\bibitem{RN015} H. Kim, A. Palacio-Morales, T. Posske, L. Rózsa, K. Palotás, L. Szunyogh, M. Thorwart, and R. Wiesendanger, Toward tailoring Majorana bound states in artificially constructed magnetic atom chains on elemental superconductors, Sci. Adv. \textbf{4}, eaar5251 (2018).

\bibitem{RN016} S. Kezilebieke, Md N. Huda, V. Vaňo, M. Aapro, S. C. Ganguli, O. J. Silveira, S. Głodzik, A. S. Foster, T. Ojanen, and P. Liljeroth, Topological superconductivity in a van der Waals heterostructure, Nature (London) \textbf{588}, 424 (2020).

\bibitem{RN017} J.-P. Xu, M.-X. Wang, Z. L. Liu, J.-F. Ge, X. Yang, C. Liu, Z. A. Xu, D. Guan, C. L. Gao, D. Qian, Y. Liu, Q.-H. Wang, F.-C. Zhang, Q.-K. Xue, and J.-F. Jia, Experimental Detection of a Majorana Mode in the core of a Magnetic Vortex inside a Topological Insulator-Superconductor ${\mathrm{Bi}}_{2}{\mathrm{Te}}_{3}/{\mathrm{NbSe}}_{2}$ Heterostructure, Phys. Rev. Lett. \textbf{114}, 017001 (2015).

\bibitem{RN018} H.-H. Sun and J.-F. Jia, Detection of Majorana zero mode in the vortex, npj Quantum Mater. \textbf{2}, 34 (2017).

\bibitem{RN019} M. Chen, X. Chen, H. Yang, Z. Du, and H.-H. Wen, Superconductivity with twofold symmetry in ${\mathrm{Bi}}_{2}{\mathrm{Te}}_{3}$/${\mathrm{Fe}}{\mathrm{Te}}_{0.55}{\mathrm{Se}}_{0.45}$ heterostructures, Science Advances, Sci. Adv. \textbf{4}, eaat1084 (2018).

\bibitem{RN020} D. Wang, L. Kong, P. Fan, H. Chen, S. Zhu, W. Liu, L. Cao, Y. Sun, S. Du, J. Schneeloch, R. Zhong, G. Gu, L. Fu, H. Ding, and H.-J. Gao, Evidence for Majorana bound states in an iron-based superconductor, Science \textbf{362}, 333 (2018).
    
\bibitem{RN021} N. Hao and J. Hu, Topological quantum states of matter in iron-based superconductors: from concept to material realization, Natl. Sci. Rev. \textbf{6}, 213 (2019).

\bibitem{RN022} Q. Liu, C. Chen, T. Zhang, R. Peng, Y.-J. Yan, C.-H.-P. Wen, X. Lou, Y.-L. Huang, J.-P. Tian, X.-L. Dong, G.-W. Wang, W.-C. Bao, Q.-H. Wang, Z.-P. Yin, Z.-X. Zhao, and D.-L. Feng, Robust and Clean Majorana Zero Mode in the Vortex Core of High-Temperature Superconductor $\mathbf{(}{\mathrm{Li}}_{0.84}{\mathrm{Fe}}_{0.16}\mathbf{)}\mathrm{OHFeSe}$, Phys. Rev. X \textbf{8}, 041056 (2018).

\bibitem{RN023} M. Li, G. Li, L. Cao, X. Zhou, X. Wang, C. Jin, C.-K. Chiu, S. J. Pennycook, Z. Wang, and H.-J. Gao, Ordered and tunable Majorana-zero-mode lattice in naturally strained LiFeAs, Nature (London) \textbf{606}, 890 (2022).

\bibitem{RN024} S. Ito, B. Feng, M. Arita, A. Takayama, R.-Y. Liu, T. Someya, W.-C. Chen, T. Iimori, H. Namatame, M. Taniguchi, C.-M. Cheng, S.-J. Tang, F. Komori, K. Kobayashi, T.-C. Chiang, and I. Matsuda, Proving Nontrivial Topology of Pure Bismuth by Quantum Confinement, Phys. Rev. Lett. \textbf{117}, 236402 (2016).

\bibitem{RN025} W. Ning, F. Kong, Y. Han, H. Du, J. Yang, M. Tian, and Y. Zhang, Robust surface state transport in thin bismuth nanoribbons, Sci. Rep. \textbf{4}, 7086 (2014).

\bibitem{RN026} W. Ning, F. Kong, C. Xi, D. Graf, H. Du, Y. Han, J. Yang, K. Yang, M. Tian, and Y. Zhang, Evidence of Topological Two-Dimensional Metallic Surface States in Thin Bismuth Nanoribbons, ACS Nano \textbf{8}, 7506 (2014).

\bibitem{RN027} S. Murakami, Quantum Spin Hall Effect and Enhanced Magnetic Response by Spin-Orbit Coupling, Phys. Rev. Lett. \textbf{97}, 236805 (2006).

\bibitem{RN028} M. Wada, S. Murakami, F. Freimuth, and G. Bihlmayer, Localized edge states in two-dimensional topological insulators: Ultrathin Bi films, Phys. Rev. B \textbf{83}, 121310 (2011).
    
\bibitem{RN029}	F. Yang, L. Miao, Z. F. Wang, M. Y. Yao, F. Zhu et al., Spatial and energy distribution of topological edge states in single Bi (111) bilayer, Phys. Rev. Lett. \textbf{109}, 016801 (2012).
     
\bibitem{RN030} S. H. Kim, K.-H. Jin, J. Park, J. S. Kim, S.-H. Jhi, T.-H. Kim, and H. W. Yeom, Edge and interfacial states in a two-dimensional topological insulator: Bi(111) bilayer on ${\mathrm{Bi}}_{2}{\mathrm{Te}}_{2}\mathrm{Se}$, Phys. Rev. B \textbf{89}, 155436 (2014).
    
\bibitem{RN031}	I. K. Drozdov, A. Alexandradinata, S. Jeon, S. Nadj-Perge, H. Ji, R. J. Cava, B. Andrei Bernevig, and A. Yazdani, One-dimensional topological edge states of bismuth bilayers, Nat. Phys. \textbf{10}, 664 (2014).
    
\bibitem{RN032}	L. Aggarwal, P. Zhu, T. L. Hughes, and V. Madhavan, Evidence for higher order topology in Bi and Bi$_{0.92}$Sb$_{0.08}$, Nat. Commun. \textbf{12}, 4420 (2021).

\bibitem{RN033} H.-H. Sun, M.-X. Wang, F. Zhu, G.-Y. Wang, H.-Y. Ma, Z.-A. Xu, Q. Liao, Y. Lu, C.-L. Gao, Y.-Y. Li, C. Liu, D. Qian, D. Guan, J.-F. Jia, Coexistence of Topological Edge State and Superconductivity in Bismuth Ultrathin Film, Nano Lett. \textbf{17}, 3035 (2017).
    
\bibitem{RN034} L. Peng, J. J. Xian, P. Z. Tang, A. Rubio, S. C. Zhang, W. H. Zhang, and Y. S. Fu, Visualizing topological edge states of single and double bilayer Bi supported on multibilayer Bi(111) films, Phys. Rev. B \textbf{98}, 245108 (2018).

\bibitem{RN035} Y. Lu, W. Xu, M. Zheng, G. Yao, L. Shen, M. Yang, Z. Luo, F. Pan, K. Wu, T. Das, J. Jiang, J. Martin, Y. P. Feng, H. Lin, and X.-S. Wang, Topological Properties Determined by Atomic Buckling in Self-Assembled Ultrathin Bi(110), Nano Lett. \textbf{15}, 80 (2015).
    
\bibitem{RN036} B. Jäck, Y. Xie, J. Li, S. Jeon, B. A. Bernevig, and A. Yazdani, Observation of a Majorana zero mode in a topologically protected edge channel, Science \textbf{364}, 1255 (2019).

\bibitem{RN037} X. Chen, M. Chen, W. Duan, H. Yang, and H.-H. Wen, Robust Zero Energy Modes on Superconducting Bismuth Islands Deposited on Fe(Te,Se), Nano Lett. \textbf{20}, 2965 (2020).

\bibitem{RN038} K. Chen, C. Wen, Z. Hou, H. Yang, and H.-H. Wen, Magnetic field evolution of the zero-energy mode in Bi islands deposited on Fe(Te,Se), Phys. Rev. B. \textbf{110}, 054510 (2024).

\bibitem{RN039} See Supplemental Material for detailed experimental methods and edge states on other islands.

\bibitem{RN040} T. Nagao, J. T. Sadowski, M. Saito, S. Yaginuma, Y. Fujikawa, T. Kogure, T. Ohno, Y. Hasegawa, S. Hasegawa, and T. Sakurai, Nanofilm allotrope and phase tansformation of ultrathin Bi film on Si(111)-$7\times7$, Phys. Rev. Lett. \textbf{93}, 105501 (2004).

\bibitem{RN041} P. Hofmann, The surfaces of bismuth: Structural and electronic properties, Prog. Surf. Sci. \textbf{81}, 191 (2006).

\bibitem{RN042} T. Hirahara, T. Nagao, I. Matsuda, G. Bihlmayer, E. V. Chulkov, Y. M. Koroteev, P. M. Echenique, M. Saito, and S. Hasegawa, Role of Spin-Orbit Coupling and Hybridization Effects in the Electronic Structure of Ultrathin Bi Films, Phys. Rev. Lett. \textbf{97}, 146803 (2006).

\bibitem{RN043} A. V. Balatsky, I. Vekhter, and J.-X. Zhu, Impurity-induced states in conventional and unconventional superconductors, Rev. Mod. Phys. \textbf{78}, 373 (2006).

\bibitem{RN044} R. Song, P. Zhang, X.-T. He, and N. Hao, Ferromagnetic impurity induced Majorana zero mode in iron-based superconductors, Phys. Rev. B \textbf{106}, L180504 (2022).    

\bibitem{RN045} A. B. Vorontsov, I. Vekhter, and M. Eschrig, Surface bound states and spin currents in noncentrosymmetric superconductors, Phys. Rev. Lett. \textbf{101}, 127003 (2008).

\bibitem{RN046} S. Sasaki, M. Kriener, K. Segawa, K. Yada, Y. Tanaka, M. Sato, and Y. Ando, Topological Superconductivity in ${\mathrm{Cu}}_{x}{\mathrm{Bi}}_{2}{\mathrm{Se}}_{3}$, Phys. Rev. Lett. \textbf{107}, 217001 (2011).
    
\bibitem{RN047} T. Mizushima, A. Yamakage, M. Sato, and Y. Tanaka, Dirac-fermion-induced parity mixing in superconducting topological insulators, Phys. Rev. B \textbf{90}, 184516 (2014).

\bibitem{RN048} J.-P. Xu, C. Liu, M.-X. Wang, J. Ge, Z.-L. Liu, X. Yang, Y. Chen, Y. Liu, Z.-A. Xu, C.-L. Gao, D. Qian, F.-C. Zhang, J.-F. Jia, Artificial Topological Superconductor by the Proximity Effect, Phys. Rev. Lett. \textbf{112}, 217001 (2014).

\bibitem{RN049} S. Wan, Q. Gu, H. Li, H. Yang, J. Schneeloch, R. D. Zhong, G. D. Gu, and H.-H. Wen, Twofold symmetry of proximity-induced superconductivity in ${\mathrm{Bi}}_{2}{\mathrm{Te}}_{3}/{\mathrm{Bi}}_{2}{\mathrm{Sr}}_{2}{\mathrm{CaCu}}_{2}{\mathrm{O}}_{8+\ensuremath{\delta}}$ heterostructures revealed by scanning tunneling microscopy, Phys. Rev. B \textbf{101}, 220503(R) (2020).

\bibitem{RN050} F. S. Bergeret, A. F. Volkov, and K. B. Efetov, Odd triplet superconductivity and related phenomena in superconductor-ferromagnet structures, Rev. Mod. Phys. \textbf{77}, 1321 (2005).
    
\bibitem{RN051} de Gennes, P. G. Boundary effects in superconductors. Rev. Mod. Phys. \textbf{36}, 225–237 (1964).

\bibitem{RN052} Y.-F. Lv, W.-L. Wang, Y.-M. Zhang, H. Ding, W. Li, L. Wang, K. He, C.-L. Song, X.-C. Ma, and Q.-K. Xue, Experimental signature of topological superconductivity and Majorana zero modes on $\beta$-Bi$_2$Pd thin films, Sci. Bull. \textbf{62}, 852 (2017).







\end{thebibliography}
\end{document}